# Tunable frequency reference by optical pumping-assisted intracavity V-type electromagnetically induced transparency

**Kang Ying [1], Yueping Niu [2,*], Dijun Chen [1,†], Haiwen Cai [1],**

**Ronghui Qu [1], Shangqing Gong [2,‡]**

[1] *Shanghai Institute of Optics and Fine Mechanics, Chinese Academy of Sciences, Shanghai 201800, China*
[2] *Department of Physics, East China University of Science and Technology, Shanghai 200237, China*
[*]*niuyp@ecust.edu.cn*
[†]*djchen@siom.ac.cn*
[‡]*sqgong@ecust.edu.cn*

**Abstract:** A tunable high resolution frequency reference is constructed using the narrowed cavity-linewidth by the optical pumping-assisted V-type electromagnetically induced transparency (EIT). At room temperature, the optical pumping effect will increase the transparency for the V-type EIT and therefore the cavity-linewidth can be narrowed apparently. For the seven EITs observed in our previous study, cavity-linewidth narrowing is observed in all of them. More importantly, we find that the cavity-linewidth can keep at 1.2*MHz* in a wide frequency range of 100*MHz* by utilizing the central EIT. This property provides a novel way for constructing high resolution tunable frequency reference via the intracavity EIT.

1.

## 1. Introduction

Electromagnetically induced transparency (EIT) is a quantum interference phenomenon occurring when two electromagnetic fields resonantly excite two different transitions sharing a common state. With the narrow transparency window in the EIT, it has been used as a frequency reference to stabilize the laser frequency. Since this idea was first demonstrated in [1], many studies have been done to construct the frequency reference via the EIT signal [2-5]. On the other hand, it is found that an EIT medium placed in an ordinary cavity can significantly narrow the cavity linewidth with large normal dispersion and almost vanishing absorption in the EIT window [7, 8], which could be a better candidate for the high resolution frequency than the EIT signal in the free space. About 1$MHz$ cavity-linewidth using hot $^{85}$Rb atoms at 87℃ [9] and cold $^{85}$Rb atoms at MOT [10] have been reported.

As we know, most of the intracavity studies have been done in the Lambda-type EIT system because the ideal Lambda-type system will produce the best EIT window compared with the V- or ladder-type system [9-18]. However, in the real experimental condition especially at room temperature, many factors would affect the EIT windows. At room temperature, the EIT effect could usually not be isolated with the optical pumping effect. In the Lambda-type EIT, the transparency would be reduced dramatically by the optical pumping effect of the strong coupling field [19]. However, in the V-type EIT case, the optical pumping effect would increase the transparency. Therefore, it may be possible to construct a high resolution frequency reference with intracavity V-type EIT at room temperature without the need for heating or trapping of the atoms.

With the detuning of the coupling field, the EIT will degrade, and hence the cavity-linewidth narrowing effect degrades. This makes the high-resolution frequency reference constructed by the EIT almost untunable. But actually, many applications require tunable high-resolution frequency references. In our previous study [20], we observed a double-peak-structured EIT in the V-type system. This EIT is a combined signal of three EITs, each of which corresponding to the case that the coupling and probe fields interact with the same ground state and excited hyperfine level. Since there is three hyperfine levels (F'=2, 3, 4) involved in the interaction, the whole effect can be considered as the coupling field interacts with the energy band formed by the three hyperfine levels. Therefore, it is expected that resonant EIT can be kept in a wide range of the coupling field frequency and hence cavity linewidth narrowing can be realized in a wide frequency range using this combined EIT.

In this paper, we do a detailed experimental investigation of intracavity EIT in V-type system at room temperature. Under suitable conditions, the linewidth is narrowed to 1.2$MHz$. This narrowing effect is equal to which using hot or cold atoms in Lambda-type EIT. Moreover, cavity linewidth narrowing at seven different cavity frequencies is observed using the seven EITs. The cavity-linewidth can keep a constant narrowed value in a wide frequency range of 100$MHz$ by utilizing the central double-peak-structure EIT, providing a novel way for high resolution tunable frequency reference via the intracavity EIT. The dependence of the narrowed cavity linewidth on the frequency detuning and coupling field power is also studied.

## 2. Experimental setup and results

The V-type EIT system used in our experiment is the $D_2$ line (780nm) of rubidium 85. A strong coupling laser and a weak probe laser are both applied between the ground state $5^2S_{1/2}$ and the excited state $5^2P_{3/2}$, as Fig. 1 shows. A diagram of the experimental setup is shown in Fig. 2. Both of the probe and coupling lasers are single-mode tunable external cavity diode lasers (ECDL) (New Focus TLB-6900), which has a linewidth of about 300 kHz. The external cavity length of the diode laser could be adjusted as tuning the control voltage to tune the laser frequency. The half-wave plate 1 (HWP1) and polarized beam splitter 1 (PB1) are used to attenuate the power of probe laser to below $50\mu W$ to avoid the saturated absorption of the $^{85}$Rb atoms and self-focusing effect. The polarized half-wave plate 2 (HWP2) and polarized beam splitter 3 (PB3) are used to attenuate the power of coupling laser to below 2mW for the V- type system to weaken the power broadening effect. The probe beam and the coupling beam (not circulating in the cavity) are focused into the cavity by lenses with focal lengths of 30 cm, and their respective beam diameters at the center of the Rb vapor cell are about $150\mu m$ and $250\mu m$. They are brought together by the polarized beam splitter 2 (PB2) and orthogonally polarized when they enter the 75-mm-long AR-coated Rb cell. The Rb vapor cell is kept in a magnetic shielding structure to eliminate the earth magnetic field effect. The reflectivity of flat mirror M1 is approximately 99.5%. The cavity mirror M2 with a reflectivity of 99.5% is concave with a 15cm radius of curvature and is controlled by a piezoelectric (PZT) driver. The finesse of the empty cavity (without the Rb cell and PB2) is 138. After we insert the Rb cell and PB2, the finesse of the cavity (with the Rb atoms off resonance) is reduced to 66 because of surface reflection losses. The free spectrum range of the empty cavity (30cm in length) is 1GHz. The coupling beam is rejected by the polarized beam splitter 4 (PB4) before reaching the detector.

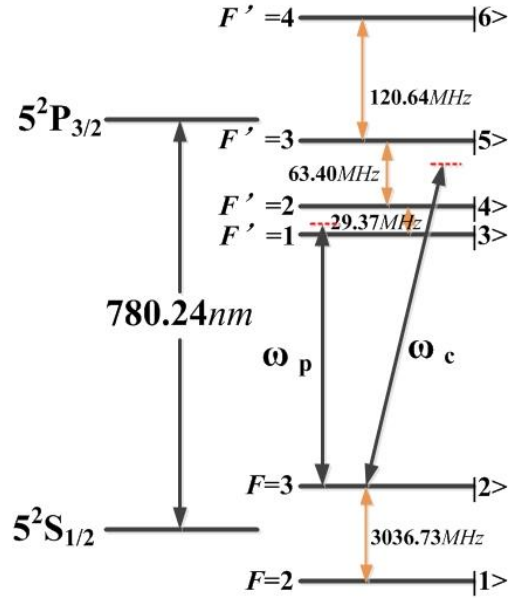

Fig. 1 The energy level configuration of the $^{85}$Rb $D_2$ line for our V-type EIT experiment

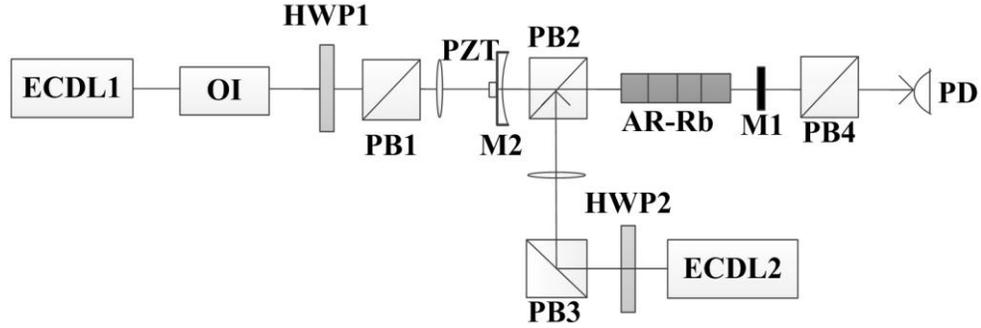

Fig. 2 Schematic diagram of the experimental setup: ECLD1 and ECLD2: probe and coupling lasers; PB1-PB4: polarizing cubic beam splitters; HWP1-HWP2: half-wave plates; PD: photodiode detector; OI: optical isolator; M1-M2: cavity mirrors.

As is known, the cavity linewidth will be affected by both the absorption and dispersion of the medium inside. Reduced absorption and enhanced dispersion would improve the narrowing effect dramatically [7]. According to the theoretical and experimental results [20], at room temperature the optical pumping effect would increase the absorption in the Lambda-type EIT window while reduce the absorption in the V-type EIT case. In our previous experiment [20], we have observed seven EIT windows including a central double-peak-structure in a V-type EIT system as we keep the probe field power at the center of the Rb cell is 45uW, as Fig. 3 shows. As described in [20], the seven EIT windows correspond to the case that there are several different probe field frequency which meet the two-photon resonance condition. While, the central peak means that the probe and coupling field interact with the same ground and excited hyperfine splitting level. Under this condition, the strong coupling field will cause Stark splitting of the two states and hence a Mollow-type absorption spectra occur. In the inset of Fig. 3, the trace is recorded as using the 0.4mW coupling field power. As shown in Fig. 3, all the seven transmission peaks are increased remarkably by the optical pumping effect when a 1.95mW strong coupling field is used. Hence, the cavity linewidth could be narrowed in different frequency using the optical pumping-assisted V-type EIT, constructing the high resolution frequency reference. For simplicity, we name the combined spectra of EIT and optical pumping effect as EIT windows in the following text.

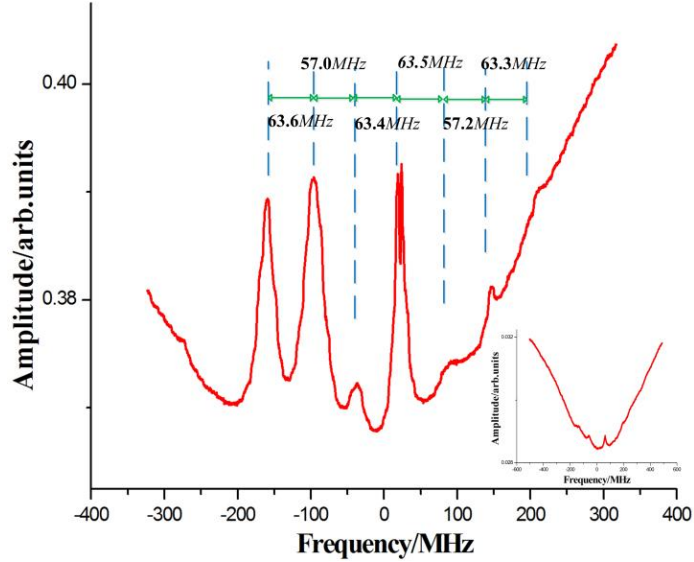

Fig. 3 Experimental trace of the seven EITs in the V-type $^{85}$Rb medium using
1.95$mW$ coupling power (the trace using 0.4$mW$ coupling power shown in inset)
The frequency axis is calibrated by the saturated apsorption spectrum

During our experiment, the coupling laser frequency is fixed (-30MHz detuning to F'=3, see the energy level in Fig. 1). By scanning the frequency of the probe laser, the transmission is recorded. To match the resonant frequency $\omega_r$ of the cavity plus medium to the probe field frequency $\omega_2$ in which the EIT occurs, the cavity length is adjusted by tuning the driving voltage of the PZT. When the condition $\omega_r = \omega_p$ is met, the transmission peak is high and narrow because of the large dispersion and the reduced absorption of the V-type EIT at room temperature. Figure 4 shows a comparison of the empty-cavity linewidth, the cavity linewidth with $^{85}$Rb vapor cell inside but the probe field off resonance and the cavity linewidth when intracavity EIT is formed. Here, the central EIT window is used and the probe and coupling field power is kept at $45\mu W$ and 1.95mW, respectively. With the coupling beam on and the probe beam tuned to the probe transition, the linewidth is 1.2MHz, which is a factor of 13 narrower than the cavity linewidth with intracavity loss but the probe field off resonance and a factor of 6 narrower than the empty cavity linewidth. In the same condition, we also record the cavity linewidth using the Lambda-type EIT. It is about 2 times wider since the optical pumping reduces the transparency in the Lambda-type EIT case at room temperature.

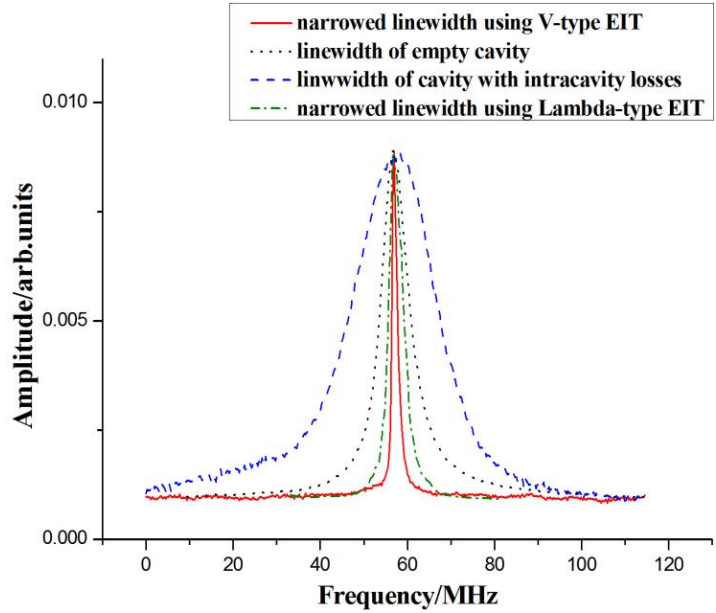

Fig. 4 Intensity of cavity output versus probe frequency, showing cavity-linewidth narrowing in different conditions.

In order to address the optical pumping effect in the V-type EIT clearly, we measure the dependence of the narrowed cavity linewidth on the intensity of the coupling beam, as shown in Fig. 5. The Rb vapor cell is kept at room temperature. The coupling field power is kept at one order of magnitude larger than the probe field power to make sure the EIT is developed in the experiment. For low coupling field power, the cavity linewidth is broad because the optical pumping effect is not fully developed and the transparency is not enhanced obviously, indicating that the pure V-type EIT haven't a good cavity linewidth narrowing effect. As the coupling field power increases, the cavity-linewidth narrows rapidly as the optical pumping-assisted EIT is developed. With the coupling field power further increasing, the linewidth slowly broadens again as an effect of power broadening. It is clearly seen that the V-type EIT reduced cavity linewidth reach a minimum value at a coupling power of approximately 2mW, just as used in our experiment. As a result, the optical pumping effect would increase the transparency in the V-type EIT case, which make the optical-pumping-assisted V-type EIT suitable for reducing cavity linewidth at room temperature.

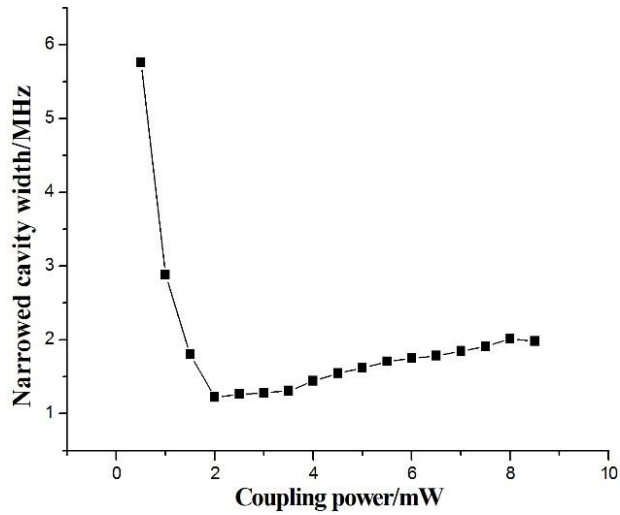

Fig. 5 Narrowed cavity linewidth verse coupling laser field power.

When we scan the cavity resonance frequency by changing the cavity length to match the probe field frequency in which the EIT occurs, we observed seven different narrowed transmittance peaks, corresponding to the seven EIT windows in the V-type system, as shown in Fig 6. The frequency separations between these narrowed-cavity peaks coincide with the frequency separations between the seven V-type EITs. So, each V-type EIT windows can be used to narrow the cavity-linewidth to construct high resolution frequency reference.

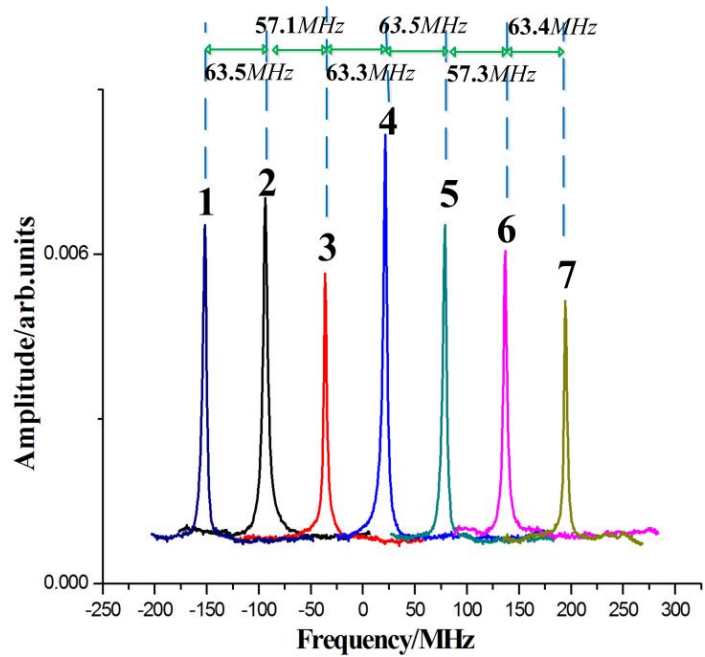

Fig. 6 Cavity linewidth narrowing in each V-type EIT window

Next, we measure each narrowed cavity linewidth under different coupling field detuning. The coupling laser frequency detuning to F'=3 from -50MHz to 50MHz with a step of 10MHz is recorded (see the energy level in Fig. 7). We define $\Delta_c = \omega_{33} - \omega_c$ with $\omega_{33}$ being the frequency separation between the level F'=3 and F=3. From Fig. 6, one can see four different tendencies of the narrowed cavity linewidth vary with the coupling field detuning, marked with circle, triangle, square and star in the curve.

According to our previous study [20], seven EIT windows appeared could be classified into four cases. In the first case, the coupling laser is considered to couple with F'=3 and two EITs occur when the probe field coupled with F'=2 and F'=4 (the second and fourth EIT in Fig. 6). When $\Delta_c$ changes from -50MHz to +50MHz, the two EITs gradually approach to the resonant condition from a far detuned condition and then far detuned again. Therefore, the cavity linewidth gradually approaches to the narrowest and then becomes wider (the triangle curves in Fig. 7). In the second case, the coupling laser is considered to couple with F'=2 and two EITs occur when the probe field coupled with F'=3 and F'=4 (the first and third EIT in Fig. 6). When $\Delta_c$ changes from -50MHz to +50MHz, the two EITs are both in a more and more detuned EIT condition. As a result, the cavity linewidth becomes wider and wider (the circle curves in Fig. 7). In the third case, the coupling laser is considered to couple with F'=4 and two EITs occur when the probe field coupled with F'=2 and F'=3 (the sixth and seventh EITs in Fig. 6). When $\Delta_c$ changes from -50MHz to +50MHz, the two EITs both gradually approach to the resonant EIT condition. As a result, the cavity linewidth is becoming narrower (the square curves in Fig. 7).

For the last one, it corresponds to the central EIT (the forth EIT window), which is a combined signal of three EITs. Each EIT comes from the probe and coupling field interact with the same ground state F=3 and excited hyperfine splitting levels F'=2, 3 and 4 [20]. It is obviously that the cavity-linewidth using this EIT keeps at about 1.2MHz in a wide tuning range of 100MHz (the star curve in Fig. 7). We think that the coupling laser could be considered as interact resonantly with the energy band formed by F'=2, F'=3 and F'=4 in this case. Thus, coupling field detuning within the range of $\Delta\omega_{F'=2->F'=4}$ is allowed for reducing the cavity linewidth. This could have potential application in the tunable high-resolution spectroscopy.

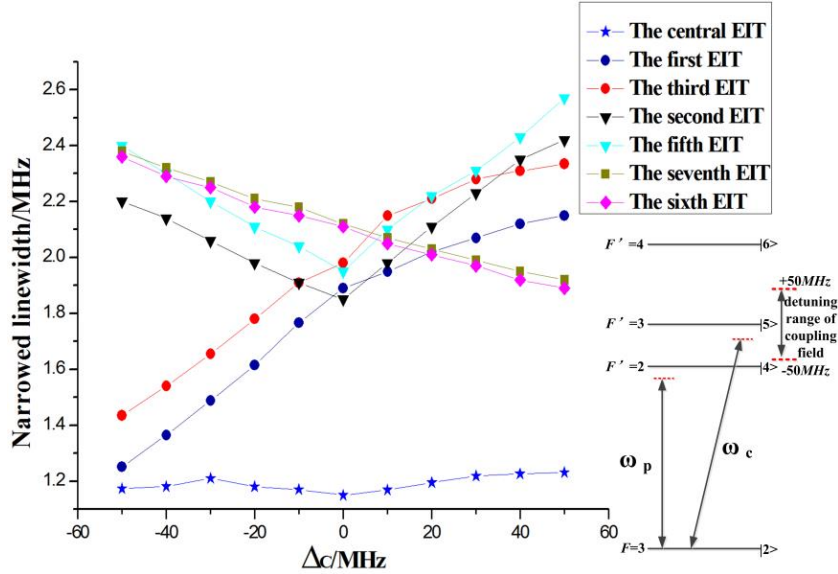

Fig. 7 Narrowed cavity linewidth in each EITs versus the detuning of the coupling laser field

## 3. Conclusion

A tunable high resolution frequency reference in a 100MHz wide frequency range is constructed using the narrowed cavity-linewidth by the optical pumping-assisted V-type EIT of the $^{85}$Rb atomic medium at room temperature. Since optical pumping reduced the absorption in the V-type EIT, the cavity-linewidth has been narrowed to 1.2MHz at room temperature. Using multi-V-type EIT windows, the cavity-linewidth narrowed in each EIT windows has been observed. A narrowed cavity transmittance signal without broadening as tuning the coupling field have been obtained using the central EIT window and this tunable high-resolution frequency reference may have potential applications in high-resolution spectroscopy and laser frequency stabilization.

**Acknowledgements:**

This work was supported by the National Natural Science Foundation of China (Grant Nos. 11274112, 61108028 and 61178031) and Shanghai Rising-Star Program of Grant No. 11QA1407400.